\documentclass[%
 reprint,%
 amssymb, amsmath,%
 aip,cha,%
]{revtex4-1}

\usepackage{docs}%
\usepackage{graphics} 
\usepackage{graphicx}
\usepackage{bm}%
\usepackage[colorlinks=true,linkcolor=blue]{hyperref}%
\expandafter\ifx\csname package@font\endcsname\relax\else
 \expandafter\expandafter
 \expandafter\usepackage
 \expandafter\expandafter
 \expandafter{\csname package@font\endcsname}%
\fi
\begin{document}

\title{Non-Twisting and Twisting Solutions of the Einstein Field Equations of a Skyrmionic String}%

\author{Malcolm Anderson$^1$,~Miftachul Hadi$^{1,2}$,~Andri Husein$^3$}%
\email{itpm.id@gmail.com (Miftachul Hadi)}
\affiliation{$^1$Department of Mathematics, Universiti Brunei Darussalam, Negara Brunei Darussalam\\
             $^2$Physics Research Centre, Indonesian Insitute of Sciences, Puspiptek, Serpong, Indonesia\\
		     $^3$Department of Physics, University of Sebelas Maret, Surakarta, Indonesia}%

\begin{abstract}
We construct non-linear sigma model plus Skyrme term (Skyrme model) with a twist in the gravitational field. We try to solve the Einstein field equations for small and large values of $r$, with and without twist. We prove that no non-twisting or twisting solutions extending from $r=0$ to $r=\infty$ exist. At last, we try to solve non-twisting and twisting solutions of the Einstein field equations with a finite radius. We find that there are no solutions with a finite radius that can satisfy the junction conditions at the boundary radius $r=r_b$, where $r_b$ is a finite radius.
\end{abstract}

\maketitle

\section{Non-Linear Sigma Model}
A non-linear sigma model is an $N$-component scalar field theory in which the fields are functions defining a mapping from the space-time to a target manifold \cite{Zakrzewski}. 
By a non-linear sigma model, we mean a field theory with the following properties \cite{hans02}:
\begin{itemize}
\item[(1)] The fields, $\phi(x)$, of the model are subject to nonlinear constraints at all points $x\in\mathcal{M}_0$, where $\mathcal{M}_0$ is the source (base) manifold, i.e. a spatial submanifold of the (2+1) or (3+1)-dimensional space-time manifold.
\item[(2)] The constraints and the Lagrangian density are invariant under the action of a global (space-independent) symmetry group, $G$, on $\phi(x)$.
\end{itemize}

The Lagrangian density of a free (without potential) nonlinear sigma model on a Minkowski background space-time is defined to be \cite{chen}
\begin{equation}\label{1}
\mathcal{L}=\frac{1}{2\lambda^2}~\gamma_{AB}(\phi)~\eta^{\mu\nu}~\partial_\mu\phi^A~\partial_\nu\phi^B
\end{equation}
where $\gamma_{AB}(\phi)$ is the field metric, $\eta^{\mu\nu}=\text{diag}(1,-1,-1,-1)$ is the Minkowski tensor, $\lambda$ is a scaling constant with dimensions of (length/energy)$^{1/2}$ and $\phi={\phi^A}$ is the collection of fields. Greek indices run from 0 to $d-1$, where $d$ is the dimension of the space-time, and upper-case Latin indices run from 1 to $N$.  

The simplest example of a nonlinear sigma model is the $O(N)$ model, which consists of $N$ real scalar fields, $\phi^A$, $\phi^B$, with the Lagrangian density \cite{hans02}
\begin{equation}\label{2}
\mathcal{L}=\frac{1}{2\lambda^2}~\delta_{AB}~\eta^{\mu\nu}~\frac{\partial\phi^A}{\partial x^\mu}~\frac{\partial\phi^B}{\partial x^\nu}
\end{equation}
where the scalar fields, $\phi^A$, $\phi^B$, satisfy the constraint
\begin{equation}\label{3}
\delta_{AB}~\phi^A\phi^B=1
\end{equation}
and $\delta_{AB}$ is the Kronecker delta. 

The Lagrangian density (\ref{2}) is obviously invariant under the global (space independent) orthogonal transformations $O(N)$, i.e. the group of $N$-dimensional rotations \cite{hans02}
\begin{equation}\label{4}
\phi^A\rightarrow\phi'^A=O^A_B~\phi^B.
\end{equation}
One of the most interesting examples of a $O(N)$ nonlinear sigma model, due to its topological properties, is the $O(3)$ nonlinear sigma model in 1+1 dimensions, with the Lagrangian density 
\begin{equation}\label{5}
\mathcal{L}=\frac{1}{2\lambda^2}~\eta^{\mu\nu}~\partial_\mu\phi~.~\partial_\nu\phi 
\end{equation}
where $\mu$ and $\nu$ range over $\{0,1\}$, and $\phi=(\phi^1,\phi^2,\phi^3)$, subject to the constraint $\phi\cdot\phi=1$, where the dot (.) denotes the standard inner product on real coordinate space of three dimensions, $R^3$. For a $O(3)$ nonlinear sigma model in any number $d$ of space-time dimensions, the target manifold is the unit sphere $S^2$ in $R^3$, and $\mu$ and $\nu$ in the Lagrangian density (\ref{5}) run from 0 to $d-1$.

A simple representation of $\phi$ (in the general time-dependent case) is
\begin{equation}\label{6}
\phi=
\begin{pmatrix}
\sin f(t,{\bf r})~\sin g(t,{\bf r}) \\
\sin f(t,{\bf r})~\cos g(t,{\bf r}) \\
\cos f(t,{\bf r})
\end{pmatrix}
\end{equation}
where $f$ and $g$ are scalar functions on the background space-time, with Minkowski coordinates $x^\mu=(t,{\bf r})$. In what follows, the space-time dimension, $d$, is taken to be 4, and so $\bf r$ is a 3-vector.

If we substitute (\ref{6}) into the Lagrangian density (\ref{5}), then it becomes 
\begin{equation}\label{7}
\mathcal{L}=\frac{1}{2\lambda^2}[\eta^{\mu\nu}~\partial_\mu f~\partial_\nu f+(\sin^2f)~\eta^{\mu\nu}~\partial_\mu g~\partial_\nu g]
\end{equation}
The Euler-Lagrange equations associated with $\mathcal{L}$ in (\ref{7}) are 
\begin{eqnarray}\label{8}
\eta^{\mu\nu}~\partial_\mu\partial_\nu f-(\sin f~\cos f)~\eta^{\mu\nu}~\partial_\mu g~\partial_\nu g=0
\end{eqnarray}
and
\begin{eqnarray}\label{9}
\eta^{\mu\nu}~\partial_\mu\partial_\nu g+2(\cot f)~\eta^{\mu\nu}~\partial_\mu f~\partial_\nu g=0.
\end{eqnarray}

\section{Soliton Solution}
Two solutions to the $O(3)$ field equations (\ref{8}) and (\ref{9}) are 
\begin{itemize}
\item[(i)] a monopole solution, which has form
\begin{equation}\label{10}
\phi=\hat{\textbf{r}}=
\begin{pmatrix}
x/\rho\\
y/\rho\\
z/\rho\\
\end{pmatrix}
\end{equation}
where $\rho=(x^2+y^2+z^2)^{1/2}$ is the spherical radius; and
\item[(ii)] a vortex solution, which is found by imposing the 2-dimensional ''hedgehog'' ansatz
\begin{equation}\label{11}
\phi=
\begin{pmatrix}
\sin f(r)~\sin (n\theta-\chi)\\
\sin f(r)~\cos (n\theta-\chi)\\
\cos f(r)
\end{pmatrix}
\end{equation}
where $r=(x^2+y^2)^{1/2}$, $\theta=\arctan (x/y)$, $n$ is a positive integer, and $\chi$ is a constant phase factor. In this thesis, we only consider the vortex solution.
\end{itemize}

A vortex is a stable time-independent solution to a set of classical field equations that has finite energy in two spatial dimensions; it is a two-dimensional soliton. In three spatial dimensions, a vortex becomes a string, a classical solution with finite energy per unit length \cite{preskill}. Solutions with finite energy, satisfying the appropriate boundary conditions, are candidate soliton solutions \cite{manton}. 

The boundary conditions that are normally imposed on the vortex solution (\ref{11}) are $f(0)=\pi$ and $\lim_{r\to\infty}f(r)=0$, so that the vortex ''unwinds'' from $\phi=-\hat{\textbf{z}}$ to $\phi=\hat{\textbf{z}}$ as $r$ increases from 0 to $\infty$. The function $f$ in this case satisfies the field equation 
\begin{equation}\label{12}
r~\frac{d^2f}{dr^2}+\frac{df}{dr}-\frac{n^2}{r}~\sin f~\cos f=0
\end{equation}
There is in fact a family of solutions to this equation (\ref{12}) satisfying the standard boundary conditions 
\begin{equation}\label{13}
\sin f=\frac{2K^{1/2}r^n}{Kr^{2n}+1}
\end{equation}
or equivalently
\begin{equation}\label{14}
\cos f=\frac{Kr^{2n}-1}{Kr^{2n}+1}
\end{equation}
where $K$ is positive constant.

The energy density, $\sigma$, of a static (time-independent) field with Lagrangian density, $\mathcal{L}$, (\ref{7}) is
\begin{eqnarray}\label{15}
\sigma 
&=& -\mathcal{L} \nonumber\\
&=& \frac{1}{2\lambda^2}\left[\eta^{\mu\nu}~\partial_\mu f~\partial_\nu f+(\sin^2 f)~\eta^{\mu\nu}~\partial_\mu g~\partial_\nu g\right]
\end{eqnarray}
The energy density of the vortex solution is
\begin{eqnarray}\label{17}
\sigma =\frac{4Kn^2}{\lambda^2}\frac{r^{2n-2}}{(Kr^{2n}+1)^2}
\end{eqnarray}
The total energy
\begin{equation}\label{18}
E=\int\int\int \sigma~ dx~dy~dz,
\end{equation}
of the vortex solution is infinite. But, the energy per unit length of the vortex solution
\begin{eqnarray}\label{19}
\mu
&=& \int\int \sigma~dx~dy=2\pi\int_0^\infty\frac{4Kn^2}{\lambda^2}\frac{r^{2n-2}}{(Kr^{2n}+1)^2}~r~dr  \nonumber\\
&=& \frac{4\pi n}{\lambda^2}
\end{eqnarray}
is finite, and does not depend on the value of $K$. (We use the same symbol for the energy per unit length and the mass per unit length, due to the equivalence of energy and mass embodied in the relation $E=mc^2$. Here, we choose units in which $c=1$). 

This last fact means that the vortex solutions in the nonlinear sigma models have no preferred scale. A small value of $K$ corresponds to a more extended vortex solution, and a larger value of $K$ corresponds to a more compact vortex solution, as can be seen by plotting $f$ (or $-\mathcal{L}$) for different values of $K$ and a fixed value of $n$. This means that the vortex solutions are what is called neutrally stable to changes in scale. As $K$ changes, the scale of the vortex changes, but the mass per unit length, $\mu$, does not. Note that because of equation (\ref{19}), there is a preferred winding number, $n=1$, corresponding to the smallest possible positive value of $\mu$.

Furthermore, it can be shown that the topological charge, $T$, of the vortex defined by
\begin{eqnarray}\label{20}
T\equiv \frac{1}{4\pi}~\varepsilon_{ABC}\int\int \phi^A~\partial_x \phi^B ~\partial_y \phi^C~dx~dy
\end{eqnarray}
where $\varepsilon_{ABC}$ is the Levi-Civita symbol, is conserved, in the sense that $\partial_t T=0$ no matter what coordinate dependence is assumed for $f$ and $g$ in (\ref{11}). 

So, the topological charge is a constant, even when the vortex solutions are perturbed. Also, it is simply shown that for the vortex solutions 
\begin{eqnarray}\label{21}
T
&=&-\frac{1}{\pi}n~[f(\infty)-f(0)]= -\frac{1}{\pi}n~(0-\pi) = n
\end{eqnarray}
and so, the winding number is just the topological charge. Because, there is no natural size for the vortex solutions, we can attempt to stabilize them by adding a Skyrme term to the Lagrangian density.  For compact twisting solutions such as the twisted baby Skyrmion string \cite{nitta1}, in addition to the topological charge, $n$, there is a second conserved quantity called the Hopf charge \cite{nitta1}, \cite{mif55}. 

\section{Skyrmion Vortex without a Twist} 
The original sigma model Lagrangian density (with the unit sphere as target manifold) is
\begin{eqnarray}\label{22}
\mathcal{L}_1=\frac{1}{2\lambda^2}~\eta^{\mu\nu}~\partial_\mu\phi~.~\partial_\nu\phi
\end{eqnarray}
If a Skyrme term is added to (\ref{22}), the result is a modified Lagrangian density
\begin{eqnarray}\label{23}
\mathcal{L}_2
&=&\frac{1}{2\lambda^2}~\eta^{\mu\nu}~\partial_\mu\phi~.~\partial_\nu\phi \nonumber\\
&&-~K_s~\eta^{\kappa\lambda}~\eta^{\mu\nu}(\partial_\kappa\phi\times\partial_\mu\phi)~.~(\partial_\lambda\phi\times\partial_\nu\phi) 
\end{eqnarray}
where the Skyrme term is the second term on the right hand side of (\ref{23}). Here, $K_s$ is a positive coupling constant.

With the choice of field representation (\ref{6}), equation (\ref{23}) becomes  
\begin{eqnarray}\label{24}
\mathcal{L}_2
&=&\frac{1}{2\lambda^2}\left(\eta^{\mu\nu}~\partial_\mu f~\partial_\nu f+\sin^2f~\eta^{\mu\nu}~\partial_\mu g~\partial_\nu g\right) \nonumber\\
&&-~K_s\left[2\sin^2f\left(\eta^{\mu\nu}~\partial_\mu f~\partial_\nu f\right)\left(\eta^{\kappa\lambda}~\partial_\kappa g~\partial_\lambda g\right) \right.\nonumber\\
&&\left.-~2\sin^2f\left(\eta^{\mu\nu}~\partial_\mu f~\partial_\nu g\right)^2\right]
\end{eqnarray}
If the vortex configuration (\ref{11}) for $\phi$ is assumed, the Lagrangian density (\ref{7}) becomes 
\begin{eqnarray}\label{25}
\mathcal{L}
&=& -\frac{1}{2\lambda^2}\left[\left(\frac{df}{dr}\right)^2 +\frac{n^2}{r^2}\sin^2f\right] \nonumber\\
&&-~2K_s\frac{n^2}{r^2}\sin^2f\left(\frac{df}{dr}\right)^2
\end{eqnarray}

The Euler-Lagrange equations generated by $\mathcal{L}_2$ (\ref{24}), namely
\begin{eqnarray}\label{26}
\partial_\alpha\left[\frac{\partial\mathcal{L}_2}{\partial(\partial_\alpha f)}\right]  -\frac{\partial\mathcal{L}_2}{\partial f}=0
\end{eqnarray}
and
\begin{eqnarray}\label{27}
\partial_\alpha\left[\frac{\partial\mathcal{L}_2}{\partial(\partial_\alpha g)}\right]  -\frac{\partial\mathcal{L}_2}{\partial g}=0
\end{eqnarray}
Reduce to a single second-order equation for $f$ 
\begin{eqnarray}\label{28}
0
&=& \frac{1}{\lambda^2}\left(\frac{d^2f}{dr^2}+\frac{1}{r}\frac{df}{dr}-\frac{n^2}{r^2}\sin f\cos f\right) \nonumber\\
&&+~4K_s~\frac{n^2}{r^2}~\sin^2f\left(\frac{d^2f}{dr^2}-\frac{1}{r}\frac{df}{dr}\right) \nonumber\\
&&+~4K_s~\frac{n^2}{r^2}~\sin f~\cos f\left(\frac{df}{dr}\right)^2
\end{eqnarray}
with the boundary conditions $f(0)=\pi$ and $\lim_{r\rightarrow\infty}f(r)=0$ as before.

If a suitable vortex solution $f(r)$ of this equation exists, it should have a series expansion for $r<<1$ of the form
\begin{eqnarray}\label{29}
f = \pi +ar+br^3+...~~\text{if}~n=1
\end{eqnarray}
or
\begin{eqnarray}\label{30}
f = \pi +ar^n+br^{3n-2}+...~~\text{if}~n\geq 2
\end{eqnarray}
where $a<0$ and $b$ are constants, and for $r>>1$ the asymptotic form
\begin{eqnarray}\label{31}
f = Ar^{-n}-\frac{1}{12}A^3r^{-3n} + ...
\end{eqnarray}
for some constant $A>0$. 

However, it turns out that it is not possible to match these small-distance and large-distance expansions if $K_s\neq 0$: meaning that any solution $f$ of (\ref{28}) either diverges at $r=0$ or as $r\rightarrow\infty$. This result follows from the following simple scaling argument.

Suppose that $f(r)$ is a solution of equation (\ref{28}). Let $q$ be any positive constant and define $f_q(r)\equiv f(qr)$. Substituting $f_q$ in place of $f$ in equation (\ref{28}) gives a value of $\mu$ which depends in general on the value of $q$
\begin{eqnarray}\label{32}
\mu_q
&=&\int\int\left\{\frac{1}{2\lambda^2}\left[\left(\frac{df_q}{dr}\right)^2+\frac{n^2}{r^2}\sin^2f_q\right] \right.\nonumber\\
&&\left.-2K_s\frac{n^2}{r^2}\sin^2f_q\left(\frac{df_q}{dr}\right)^2\right\}r~dr~d\theta
\end{eqnarray}
where
\begin{eqnarray}\label{33}
\frac{df_q}{dr} =qf'(qr)
\end{eqnarray}

So, if $r$ is replaced as the variable of integration by $\overline{r}=qr$, we have
\begin{eqnarray}\label{34}
\mu_q
&=&\int\int\left\{\frac{1}{2\lambda^2}\left[\left(\frac{df(\overline{r})}{d\overline{r}}\right)^2+\frac{n^2}{\overline{r}^2}\sin^2f(\overline{r})\right] \right.\nonumber\\
&&\left.+~2q^2K_s\frac{n^2}{\overline{r}^2}\sin^2f(\overline{r})\left(\frac{df(\overline{r})}{d\overline{r}}\right)^2\right\}\overline{r}~d\overline{r}~d\theta
\end{eqnarray}
In particular,
\begin{eqnarray}\label{35}
\left.\frac{\partial\mu_q}{\partial q}\right|_{q=1}
&=& 4qK_s\int\int\frac{n^2}{\overline{r}^2}\sin^2f(\overline{r})\left(\frac{df(\overline{r})}{d\overline{r}}\right)^2\overline{r}d\overline{r}d\theta\nonumber\\
&>& 0
\end{eqnarray}
But, if $f$ is a localized solution of eq.(\ref{28}), meaning that it remains suitably bounded as $r\rightarrow 0$ and as $r\rightarrow\infty$, it should be a stationary point of $\mu$, meaning that $\partial\mu_q/\partial q|_{q=1}=0$. 

It follows therefore that no localized solution of (\ref{28}) exists. A more rigorous statement of this property follows on from Derrick's theorem \cite{derrick}, which states that a necessary condition for vortex stability is that 
\begin{eqnarray}\label{36}
\left.\frac{\partial \mu}{\partial q}\right|_{q=1}&=&0
\end{eqnarray}
It is evident that (\ref{35}) does not satisfy this criterion.

In an attempt to fix this problem, we could add a ''mass'' term i.e. $K_v(1-\hat{\textbf{z}}.\phi)$, to the Lagrangian density, $\mathcal{L}_2$, where $\hat{\textbf{z}}$ is the direction of $\phi$ at $r=\infty$ (where $f(r)=0$). The Lagrangian density then becomes
\begin{eqnarray}\label{37}
\mathcal{L}_3=\mathcal{L}_2+K_v(1-\underline{n}.\hat{\underline{\phi}})
\end{eqnarray}
[This Lagrangian density corresponds to the baby Skyrmion model in equation (2.2), p.207 of \cite{piette1}]. 

The kinetic term (in the case of a free particle) together with the Skyrme term in $\mathcal{L}_2$ are not sufficient to stabilize a baby Skyrmion, as the kinetic term in $2+1$ dimensions is conformally (scale) invariant and the baby Skyrmion can always reduce its energy by inflating indefinitely. This is in contrast to the usual Skyrme model, in which the Skyrme term prohibits the collapse of the $3+1$ soliton \cite{gisiger}. The mass term is added to limit the size of the baby Skyrmion. 

\section{Skyrmion Vortex with a Twist}
Instead of adding a mass term to stabilize the vortex, we will retain the baby Skyrme model Lagrangian (\ref{24}) but include a twist in the field, $g$, in (\ref{11}). That is, instead of choosing \cite{simanek}, \cite{cho}
\begin{equation}\label{38}
g=n\theta-\chi
\end{equation}
we choose
\begin{equation}\label{39}
g=n\theta+mkz
\end{equation}
where $mkz$ is the twist term, $m$ and $n$ are integers, $2\pi/k$ is the period in the $z$-direction.

The Lagrangian density (\ref{24}) then becomes 
\begin{eqnarray}\label{40}
\mathcal{L}_2
&=& \frac{1}{2\lambda^2}\left[\left(\frac{df}{dr}\right)^2+\sin^2f\left(\frac{n^2}{r^2}+m^2k^2\right)\right] \nonumber\\
&&+~2K_s\sin^2f\left(\frac{df}{dr}\right)^2\left(\frac{n^2}{r^2}+m^2k^2\right)
\end{eqnarray}
The value of the twist lies in the fact that in the far field, where $r\to\infty$ then $f\to0$, the Euler-Lagrange equations for $f$ for both $\mathcal{L}_3$ (without a twist) and $\mathcal{L}_2$ (with a twist) are formally identical to leading order, with $m^2k^2/\lambda^2$ in the twisted case playing the role of the mass coupling constant, $K_v$. So, it is expected that the twist term will act to stabilize the vortex just as the mass term does in $\mathcal{L}_3$.

On a physical level, the twist can be identified with a circular stress in the plane, perpendicular to the vortex string (which can be imagined e.g. as a rod aligned with the $z$-axis). The direction of the twist can be clockwise or counter-clockwise. In view of the energy-mass relation, the energy embodied in the stress term contributes to the gravitational field of the string, with the net result that the trajectories of freely-moving test particles differ according to whether they are directed clockwise or counter-clockwise around the string.

The Euler-Lagrange equation corresponding to the twisted Skyrmion string Lagrangian density (\ref{40}) reads
\begin{eqnarray}\label{41}
0
&=&\frac{1}{\lambda^2}\left[\frac{d^2f}{dr^2}+\frac{1}{r}~\frac{df}{dr}-\left(\frac{n^2}{r^2}+m^2k^2\right)~\sin f~\cos f\right] \nonumber\\
&&+~4K_s~\frac{n^2}{r^2}~\sin^2f\left(\frac{d^2f}{dr^2} -\frac{1}{r}\frac{df}{dr}\right) \nonumber\\
&&+~4K_s~m^2k^2\sin^2f\left(\frac{d^2f}{dr^2}+\frac{1}{r}\frac{df}{dr}\right)\nonumber\\
&&+~4\left(\frac{n^2}{r^2}+m^2k^2\right)K_s~\sin f~\cos f\left(\frac{df}{dr}\right)^2
\end{eqnarray}
It should be noted that the second Euler-Lagrange equation (\ref{27}) is satisfied identically if $g$ has the functional form (\ref{39}).

\section{The Gravitational Field of a Twisted Skyrmion String}
We are interested in constructing the space-time generated by a twisted Skyrmion string. Without gravity, the Lagrangian density of the system is $\mathcal{L}_2$, as given in equation (\ref{24}).
To add gravity, we replace $\eta^{\mu\nu}$ in $\mathcal{L}_2$ with a space-time metric tensor, $g^{\mu\nu}$, which in view of the time-independence and cylindrical symmetry of the assumed vortex solution is taken to be a function of $r$ alone. 

The contravariant metric tensor, $g^{\mu\nu}$, is of course the inverse of the covariant metric tensor, $g_{\mu\nu}$, of the space-time meaning that $g^{\mu\nu}=(g_{\mu\nu})^{-1}$. We use a cylindrical coordinate system $(t,r,\theta,z)$, where $t$ and $z$ have unbounded range, $r\in[0,\infty)$ and $\theta\in[0,2\pi)$. 

The components of the metric tensor
\begin{eqnarray}\label{74}
g_{\mu\nu}
=
\begin{pmatrix}
g_{tt} & 0      & 0                & 0  \\
0      & g_{rr} & 0                & 0  \\
0      & 0      & g_{\theta\theta} & g_{\theta z} \\
0      & 0     & g_{z\theta}      & g_{zz} \\
\end{pmatrix}
\end{eqnarray}
are all functions of $r$, and the presence of the off-diagonal components $g_{\theta z}=g_{z\theta}$ reflects the twist in the space-time (see the twist term, $mkz$, in previous equation i.e. $g=n\theta +mkz$).

The Lagrangian we will be using is
\begin{eqnarray}\label{75}
\mathcal{L}_4
&=&\frac{1}{2\lambda^2}(g^{\mu\nu}~\partial_\mu f~\partial_\nu f+\sin^2f~ g^{\mu\nu}~\partial_\mu g~\partial_\nu g) \nonumber\\
&&-~2K_s\sin^2f~[(g^{\mu\nu}~\partial_\mu f~\partial_\nu f)(g^{\kappa\lambda}~\partial_\kappa g~\partial_\lambda g) \nonumber\\
&&-~2\sin^2f~(g^{\mu\nu}~\partial_\mu f~\partial_\nu g)^2]
\end{eqnarray}
where $f=f(r)$ and $g=n\theta+mkz$.

We need to solve:
\begin{itemize}
\item[(i)] the Einstein equations
\begin{eqnarray}\label{76}
G_{\mu\nu}
&=& -\frac{8\pi G}{c^4}~T_{\mu\nu}  
\end{eqnarray}
where the stress-energy tensor of the vortex, $T_{\mu\nu}$, is defined by
\begin{eqnarray}\label{77}
T_{\mu\nu}
&\equiv&2\frac{\partial\mathcal{L}_4}{\partial g^{\mu\nu}}-g_{\mu\nu}~\mathcal{L}_4
\end{eqnarray}
and
\begin{eqnarray}\label{78}
G_{\mu\nu} 
&=& R_{\mu\nu}-\frac{1}{2}g_{\mu\nu}~R 
\end{eqnarray}
with
$R^{\mu\nu}$ the Ricci tensor and 
\begin{eqnarray}\label{79}
R=g_{\mu\nu}~R^{\mu\nu} = g^{\mu\nu}~R_{\mu\nu}
\end{eqnarray}
the Ricci scalar; and
\item[(ii)] the field equations for $f$ and $g$
\begin{eqnarray}\label{80}
\nabla^\mu\frac{\partial\mathcal{L}_4}{\partial(\partial f/\partial x^\mu)}=\frac{\partial\mathcal{L}_4}{\partial f};~~\nabla^\mu\frac{\partial\mathcal{L}_4}{\partial(\partial g/\partial x^\mu)}=\frac{\partial\mathcal{L}_4}{\partial g}
\end{eqnarray}
\end{itemize}

However, the field equations for $f$ and $g$ are in fact redundant, as they are satisfied identically whenever the Einstein equations are satisfied, by virtue of the Bianchi identities (i.e. permuting of the covariant derivative of the Riemann tensor) $\nabla_\mu G^{\mu}_{\nu}=0$. So, only the Einstein equations will be considered in this section.

To simplify the Einstein equations, we first choose a gauge condition that narrows down the form of the metric tensor. The gauge condition preferred here is that
\begin{eqnarray}\label{81}
g_{\theta\theta}~g_{zz}-(g_{\theta z})^2=r^2
\end{eqnarray}
The geometric significance of this choice is that the determinant of the 2-metric tensor projected onto the surfaces of constant $t$ and $z$ is $r^2$, and so the area element on these surfaces is just $r~dr~d\theta$.

As a further simplification, we write
\begin{eqnarray}\label{82}
g_{tt}=A^2;~~g_{rr}=-B^2;~~g_{\theta\theta}=-C^2;~~g_{\theta z}=\omega
\end{eqnarray}
where $A(r)$, $B(r)$, $C(r)$, $\omega(r)$ and so
\begin{eqnarray}\label{83}
g_{zz}=-\left(\frac{r^2+\omega^2}{C^2}\right).
\end{eqnarray}
The metric tensor, $g_{\mu\nu}$, therefore has the form
\begin{eqnarray}\label{84}
g_{\mu\nu}=
\begin{pmatrix}
A^2  &  0     &    0     &  0  \\
0    &  -B^2  &    0     &  0  \\
0    &  0     &  -C^2    &  \omega  \\
0    &  0     &   \omega & -\left(\frac{r^2+\omega^2}{C^2}\right)
\end{pmatrix}
\end{eqnarray}

\section{Einstein Field Equations}
The Einstein tensor, $G_{\mu\nu}$, is defined as
\begin{equation}\label{210}
G_{\mu\nu} \equiv R_{\mu\nu} - \frac{1}{2}g_{\mu\nu}~R
\end{equation}
where $R_{\mu\nu}$ is the Ricci curvature tensor, $R$ is the Ricci scalar and $g_{\mu\nu}$ is the metric tensor. 

Using (\ref{79}) then (\ref{210}) can be rewritten as 
\begin{eqnarray}\label{210.1}
G_{\mu\nu}\nonumber
&=& R_{\mu\nu} - \frac{1}{2}g_{\mu\nu}~R = R_{\mu\nu} - \frac{1}{2}g_{\mu\nu}~g^{\alpha\beta}~R_{\alpha\beta}\nonumber\\
&=& \delta^{\alpha}_{\mu}~\delta^{\beta}_{\nu}~R_{\alpha\beta} - \frac{1}{2}g_{\mu\nu}~g^{\alpha\beta}~R_{\alpha\beta} \nonumber\\
&=& \big(\delta^{\alpha}_{\mu}~\delta^{\beta}_{\nu} - \frac{1}{2}g_{\mu\nu}~g^{\alpha\beta}\big)R_{\alpha\beta}\nonumber\\
&=& \big(\delta^{\alpha}_{\mu}~\delta^{\beta}_{\nu} - \frac{1}{2}g_{\mu\nu}~g^{\alpha\beta}\big)\nonumber\\
&&\times\big(\Gamma^{\rho}_{\alpha\beta,\rho}-\Gamma^{\rho}_{\alpha\rho,\beta}+\Gamma^{\rho}_{\rho\lambda}\Gamma^{\lambda}_{\alpha\beta}-\Gamma^{\rho}_{\beta\lambda}\Gamma^{\lambda}_{\rho\alpha}\big)
\end{eqnarray}
where $\delta^{\alpha}_{\mu}$, $\delta^{\beta}_{\nu}$ are Kronecker deltas,
\begin{eqnarray}\label{210.2}
\delta^{\alpha}_{\mu}
=\Big\{
\begin{matrix}
1~~~~~\text{if $\alpha = \mu$,} \\
0~~~~~\text{if $\alpha \not= \mu$,} \\
\end{matrix}~~~~~~~~~~
\delta^{\beta}_{\nu}
=\Big\{
\begin{matrix}
1~~~~~\text{if $\beta = \nu$} \\
0~~~~~\text{if $\beta \not= \nu$} \\
\end{matrix}
\end{eqnarray}

We now calculate (\ref{210}) using covariant metric tensor below
\begin{eqnarray}\label{210.4}
g_{tt}
&=&A(r)^2;~~~g_{rr}=-B(r)^2;~~~g_{\theta\theta} = -C(r)^2 \nonumber\\
g_{\theta z}&=&g_{z\theta}=\omega (r);~~~g_{zz}=-\left(\frac{r^2+\omega^2}{C^2}\right)
\end{eqnarray}
From (\ref{210.1}) we obtain 
\begin{eqnarray}\label{210.5}
G_{tt}
&=& \frac{A^2C'^2}{B^2C^2}\Big(1+\frac{\omega^2}{r^2}\Big) - \frac{A^2B'}{rB^3} - \frac{\omega A^2\omega'C'}{r^2B^2C} \nonumber\\
&&-~\frac{A^2C'}{rB^2C} + \frac{A^2\omega'^2}{4r^2B^2} 
\end{eqnarray}
\begin{eqnarray}\label{211}
G_{rr}
&=& \frac{C'^2}{C^2}\Big(1+\frac{\omega^2}{r^2}\Big) -\frac{\omega\omega'C'}{r^2C} -\frac{C'}{rC} -\frac{A'}{rA} 
+\frac{\omega'^2}{4r^2}
\end{eqnarray}
\begin{eqnarray}\label{212}
G_{\theta\theta}
&=& R_{\theta\theta} -\frac{1}{2}g_{\theta\theta}~R = R_{\theta\theta} -\frac{1}{2}(-C^2)~R \nonumber\\
&=& R_{\theta\theta} + \frac{C^2R}{2}
\end{eqnarray}
Substituting $R_{\theta\theta}$ and $R$ into (\ref{212}), we obtain
\begin{eqnarray}\label{213}
G_{\theta\theta}
&=& \frac{C^2}{B^2}\Big[\frac{C''}{C} - \frac{A''}{A} + \frac{A'C'}{AC} - \frac{B'C'}{BC} + \frac{A'B'}{AB}  - \frac{A'}{rA} \nonumber\\
&& +\frac{C'^2}{C^2} - \frac{2B'}{rB} - \frac{C'}{rC}\Big] - \frac{3C^2}{B^2}\Big[\frac{C'^2}{C^2}\Big(1+\frac{\omega^2}{r^2}\Big) \nonumber\\
&& - \frac{\omega \omega' C'}{r^2C} - \frac{C'}{rC} - \frac{B'}{rB} + \frac{\omega'^2}{4r^2}\Big]
\end{eqnarray}
\begin{eqnarray}\label{214}
G_{\theta z}
&=& R_{\theta z} - \frac{\omega R}{2}
\end{eqnarray}
Substituting $R_{\theta z}$ and $R$ into (\ref{214}), we obtain 
\begin{eqnarray}\label{215}
G_{\theta z}
&=& \frac{\omega}{B^2}\Big[\frac{A''}{A} - \frac{\omega''}{2\omega} - \frac{A'\omega'}{2\omega A} + \frac{B'\omega'}{2\omega B} - \frac{A'B'}{AB} + \frac{A'}{rA} \nonumber\\
&& +\frac{\omega'}{2\omega r} +\frac{2B'}{rB}\Big] +\frac{3\omega}{B^2}\Big[ \frac{C'^2}{C^2}\Big(1+\frac{\omega^2}{r^2}\Big) - \frac{\omega\omega'C'}{r^2C} \nonumber\\
&& - \frac{C'}{rC} - \frac{B'}{rB} + \frac{\omega'^2}{4r^2} \Big]
\end{eqnarray}
\begin{eqnarray}\label{216}
G_{zz}
&=& -\frac{\omega^2A'}{rAB^2C^2} + \frac{\omega^2 B'}{rB^3C^2} + \frac{4\omega^2C'}{rB^2C^3} - \frac{\omega\omega'}{rB^2C^2} \nonumber\\
&& + \frac{\omega\omega' A'}{AB^2C^2} - \frac{\omega\omega' B'}{B^3C^2}  +\frac{\omega\omega'C'}{B^2C^3}\Big(\frac{3\omega^2}{r^2}-1\Big)  \nonumber\\
&& + \frac{\omega'^2}{4B^2C^2}\Big(1-\frac{3\omega^2}{r^2}\Big) + \frac{r^2A'B'}{AB^3C^2}\Big(1+\frac{\omega^2}{r^2}\Big)\nonumber\\
&& -\frac{r^2A'C'}{AB^2C^3}\Big(1+\frac{\omega^2}{r^2}\Big) + \frac{r^2B'C'}{B^3C^3}\Big(1+\frac{\omega^2}{r^2}\Big) \nonumber\\
&& - \frac{r^2A''}{AB^2C^2}\Big(1+\frac{\omega^2}{r^2}\Big)  - \frac{r^2C''}{B^2C^3}\Big(1+\frac{\omega^2}{r^2}\Big) \nonumber\\
&& + \frac{\omega\omega''}{B^2C^2} - \frac{3\omega^2C'^2}{B^2C^4}\Big(1+\frac{\omega^2}{r^2}\Big)
\end{eqnarray}
Multiplying (\ref{210.5}) with $B^2/A^2$, we obtain
\begin{eqnarray}\label{217}
\frac{B^2}{A^2}G_{tt}
&=& \frac{C'^2}{C^2}\Big(1+\frac{\omega^2}{r^2}\Big)  - \frac{\omega \omega'C'}{r^2C}-\frac{C'}{rC} \nonumber\\
&& - \frac{B'}{rB} + \frac{\omega'^2}{4r^2}
\end{eqnarray}
Comparing this equation with (\ref{211}), we see that
\begin{eqnarray}\label{218}
G_{rr}
&=& \frac{B'}{rB} -\frac{A'}{rA} +\frac{B^2}{A^2}G_{tt}
\end{eqnarray}
From (\ref{213}) we obtain
\begin{eqnarray}\label{219}
\frac{B^2}{C^2}G_{\theta\theta}
&=& \frac{C''}{C} - \frac{A''}{A} + V' - \frac{3B^2}{A^2}G_{tt}
\end{eqnarray}
where
\begin{eqnarray}\label{220}
V'
&=& \frac{A'C'}{AC} - \frac{B'C'}{BC} + \frac{A'B'}{AB}  - \frac{A'}{rA} +\frac{C'^2}{C^2} \nonumber\\
&& - \frac{2B'}{rB} - \frac{C'}{rC}
\end{eqnarray}
From (\ref{215}) we obtain
\begin{eqnarray}\label{221}
\frac{B^2}{\omega}G_{\theta z}
&=& \frac{A''}{A} - \frac{\omega''}{2\omega} + W' +\frac{3B^2}{A^2}G_{tt}
\end{eqnarray}
where
\begin{eqnarray}\label{222}
W'
&=& - \frac{A'\omega'}{2\omega A} + \frac{B'\omega'}{2\omega B} - \frac{A'B'}{AB} + \frac{A'}{rA} + \frac{\omega'}{2\omega r} \nonumber\\
&& +\frac{2B'}{rB}
\end{eqnarray}
From (\ref{216}) we obtain
\begin{eqnarray}\label{223}
\frac{B^2C^2}{\omega^2}G_{zz}
&=& -\frac{A''}{A}\Big(1+\frac{r^2}{\omega^2}\Big) -\frac{C''}{C}\Big(1+\frac{r^2}{\omega^2}\Big) \nonumber\\
&& +\frac{\omega''}{\omega} +X'  -\frac{3B^2}{A^2}G_{tt}
\end{eqnarray}
where
\begin{eqnarray}\label{224}
X'
&=&  -~\frac{A'}{rA} - \frac{\omega'}{\omega r}+ \frac{\omega'A'}{\omega A} - \frac{\omega'B'}{\omega B} + \frac{A'B'}{AB}\Big(1+\frac{r^2}{\omega^2}\Big) \nonumber\\
&& - \frac{A'C'}{AC}\Big(1+\frac{r^2}{\omega^2}\Big) +\frac{B'C'}{BC}\Big(1+\frac{r^2}{\omega^2}\Big) - \frac{2B'}{rB} \nonumber\\
&& - \frac{\omega'C'}{\omega C} + \frac{C'}{rC} + \frac{\omega'^2}{4\omega^2}
\end{eqnarray}

The Einstein field equations can be defined in the form
\begin{eqnarray}\label{225}
G_{\mu\nu}
&\equiv & -\varepsilon T_{\mu\nu}
\end{eqnarray}
where $\varepsilon = \frac{8\pi G}{c^4}$. 

From (\ref{225}), we obtain
\begin{equation}\label{226}
T_{tt} = \frac{A^2}{2B^2}\Big[\Big(\frac{1}{\lambda^2} + 4NK_s\sin^2f\Big)\left(\frac{\partial f}{\partial r}\right)^2 - \frac{A^2N\sin^2f}{2\lambda^2}\Big]
\end{equation}
\begin{eqnarray}\label{227}
T_{rr} 
&=& \frac{B^2}{A^2}T_{tt} + \frac{B^2N\sin^2f}{\lambda^2} 
\end{eqnarray}
\begin{eqnarray}\label{228}
T_{\theta\theta} 
&=& -\frac{C^2}{A^2}T_{tt} +\left[\frac{1}{\lambda^2} -\frac{4K_s}{B^2}\left(\frac{\partial f}{\partial r}\right)^2\right]n^2\sin^2f
\end{eqnarray}
\begin{eqnarray}\label{229}
T_{\theta z} 
&=& \frac{\omega}{A^2}T_{tt} +\left[\frac{1}{\lambda^2} -\frac{4K_s}{B^2}\left(\frac{\partial f}{\partial r}\right)^2\right]nmk~\sin^2f
\end{eqnarray}
\begin{eqnarray}\label{230}
T_{zz} 
&=& -\frac{r^2+\omega^2}{A^2C^2}T_{tt} +\left[\frac{1}{\lambda^2} -\frac{4K_s}{B^2}\left(\frac{\partial f}{\partial r}\right)^2\right]m^2k^2\sin^2f \nonumber\\
\end{eqnarray}
Eliminating $A''$ from eqs.(\ref{219}) and (\ref{213}) we obtain
\begin{eqnarray}\label{236}
\frac{B^2}{C^2}G_{\theta\theta} + \frac{B^2}{\omega}G_{\theta z} 
&=& \frac{C''}{C} - \frac{\omega''}{2\omega} + V' + W'
\end{eqnarray}
Eliminating $A''$ from eqs.(\ref{213}) and (\ref{223}), we obtain
\begin{eqnarray}\label{237}
&&
\frac{B^2}{\omega}\Big(1+\frac{r^2}{\omega^2}\Big)G_{\theta z} + \frac{B^2C^2}{\omega^2}G_{zz} \nonumber\\
&=& -~\frac{C''}{C}\Big(1+\frac{r^2}{\omega^2}\Big) + \frac{\omega''}{2\omega}\Big(1-\frac{r^2}{\omega^2}\Big)\nonumber\\
&& +~ W'\Big(1+\frac{r^2}{\omega^2}\Big) + X' + \frac{3B^2r^2}{\omega^2A^2}G_{tt}
\end{eqnarray}
Eliminating $C''$ from eqs.(\ref{236}) and (\ref{237}), we obtain 
\begin{eqnarray}\label{238}
\omega'' 
&=& \varepsilon\omega\Big(\frac{B^2}{\lambda^2} - 4K_sf'^2\Big)\Big[\Big(\frac{n^2}{C^2} +  \frac{2nmk}{\omega}\Big)\Big(1+\frac{\omega^2}{r^2}\Big) \nonumber\\
&& +\frac{C^2m^2k^2}{r^2}\Big]\sin^2(f)  +\frac{4\omega C'^2}{C^2}\Big(1+\frac{\omega^2}{r^2}\Big)-\frac{4\omega^2\omega'C'}{r^2C} \nonumber\\
&& -\frac{4\omega C'}{rC} +\frac{\omega\omega'^2}{r^2} +\frac{\omega A'}{rA} - \frac{\omega B'}{rB}  \nonumber\\
&& -\frac{A'\omega'}{A} +\frac{B'\omega'}{B}+\frac{\omega'}{r}
\end{eqnarray}

Substituting eq.(\ref{238}) into (\ref{237}), we obtain 
\begin{eqnarray}\label{239}
C'' 
&=& \frac{\varepsilon C}{2}\Big(\frac{B^2}{\lambda^2}-4K_s f'^2\Big)\Big[\frac{n^2}{C^2}\Big(\frac{\omega^2}{r^2}-1\Big)+\frac{2nmk\omega}{r^2} \nonumber\\
&& + \frac{C^2m^2k^2}{r^2}\Big]\sin^2(f) +\frac{C'^2}{C}\Big(1+\frac{2\omega^2}{r^2}\Big) - \frac{2\omega\omega'C'}{r^2} \nonumber\\
&& -\frac{C'}{r} - \frac{CB'}{2rB} + \frac{C\omega'^2}{2r^2} - \frac{A'C'}{A} + \frac{B'C'}{B} + \frac{CA'}{2rA} \nonumber\\
\end{eqnarray}
Substituting eq.(\ref{239}) into (\ref{236}), we obtain 
\begin{eqnarray}\label{240}
A''
&=& \frac{\varepsilon A}{2}\Big(\frac{B^2}{\lambda^2} - 4K_sf'^2\Big)\Big[\frac{n^2}{C^2}\Big(1+\frac{\omega^2}{r^2}\Big)+\frac{2nmk\omega}{r^2} \nonumber\\
&& +\frac{C^2m^2k^2}{r^2}\Big]\sin^2(f)  +\frac{A'B'}{B} - \frac{A'}{2r} - \frac{AB'}{2rB} 
\end{eqnarray}
Using (\ref{218}), we obtain
\begin{eqnarray}\label{242}
B'
&=& \frac{BA'}{A} + \frac{r\varepsilon B^3}{\lambda^2}\Big[\frac{n^2}{C^2}\Big(1+\frac{\omega^2}{r^2}\Big)  + \frac{2nmk\omega}{r^2} \nonumber\\
&&+\frac{C^2m^2k^2}{r^2}\Big]\sin^2f
\end{eqnarray}
where
\begin{equation}\label{243}
Nr^2 =  -\frac{n^2(r^2+\omega^2)}{C^2} - 2nmk\omega - C^2m^2k^2
\end{equation}
Using (\ref{225}), (\ref{226}) we obtain
\begin{eqnarray}\label{245}
f'
&=&  \Big(\frac{\varepsilon}{2\lambda^2}+2\varepsilon NK_s\sin^2f\Big)^{-1/2}\nonumber\\
&&\times~\Big[\frac{\varepsilon A^2N\sin^2f}{4\lambda^2} -\frac{C'^2}{C^2}\Big(1+\frac{\omega^2}{r^2}\Big) 
\nonumber\\
&&+\frac{\omega\omega'C'}{r^2C} +\frac{C'}{rC} +\frac{B'}{rB} -\frac{\omega'^2}{4r^2}\Big]^{1/2}
\end{eqnarray}
Using (\ref{242}) to eliminate $B'$, eq.(\ref{245}) can be rewritten as
\begin{eqnarray}\label{247}
f'
&=& \left\{\frac{\varepsilon}{2\lambda^2} -2\varepsilon K_s\Big[\frac{n^2}{C^2}\Big(1+\frac{\omega^2}{r^2}\Big) +\frac{2nmk\omega}{r^2} 
\right.\nonumber\\
&&\left. +\frac{C^2m^2k^2}{r^2}\Big]\sin^2f\right\}^{-1/2}\nonumber\\
&& \times~\left\{\frac{A'}{rA}+ \frac{\omega\omega'C'}{r^2C} + \frac{C'}{rC} -\frac{C'^2}{C^2}\Big(1+\frac{\omega^2}{r^2}\Big)  
-\frac{\omega'^2}{4r^2} \right.\nonumber\\
&&\left. +\frac{\varepsilon A^2}{4\lambda^2}
\left[\frac{n^2}{C^2}\left(1+\frac{\omega^2}{r^2}\right) +\frac{2nmk\omega}{r^2} +\frac{C^2m^2k^2}{r^2}\right]\sin^2f\right\}^{1/2} \nonumber\\
\end{eqnarray}
Using (\ref{243}), eq.(\ref{240}) can be rewritten as 
\begin{eqnarray}\label{248}
A'' 
&=& -\frac{A'}{r} +\frac{A'B'}{B} +2\varepsilon K_s AN\sin^2ff'^2
\end{eqnarray}
or
\begin{eqnarray}\label{249}
A'' 
&=& -\frac{A'}{r} + \frac{A'^2}{A} +\varepsilon\sin^2f\Big[\frac{n^2}{C^2}\Big(1+\frac{\omega^2}{r^2}\Big) \nonumber\\
&&+\frac{2nmk\omega}{r^2} + \frac{C^2m^2k^2}{r^2}\Big]\nonumber\\
&& \times \left(\frac{rA'B^2}{\lambda^2} - 2K_s\left\{\frac{A'}{r}+ \frac{\omega A\omega'C'}{r^2C}  \right.\right.\nonumber\\
&&\left.\left.+~\frac{AC'}{rC}  -\frac{AC'^2}{C^2}\left(1+\frac{\omega^2}{r^2}\right)  -\frac{A\omega'^2}{4r^2} \right.\right.\nonumber\\
&&\left.\left. +\frac{\varepsilon AB^2}{2\lambda^2}\Big[\frac{n^2}{C^2}\left(1+\frac{\omega^2}{r^2}\right) +\frac{2nmk\omega}{r^2} 
+\frac{C^2m^2k^2}{r^2}\Big]\right.\right.\nonumber\\
&&\left.\left.\times \sin^2f\right\}\right.\nonumber\\
&&\left.\times \left\{\frac{\varepsilon}{2\lambda^2} - 2\varepsilon K_s\Big[\frac{n^2}{C^2}\Big(1+\frac{\omega^2}{r^2}\Big) 
+\frac{2nmk\omega}{r^2}  \right.\right.\nonumber\\
&&\left.\left.+\frac{C^2m^2k^2}{r^2}\Big]\sin^2f\right\}^{-1}\right)
\end{eqnarray}

In order to separate the equations for the purposes of numerical integration, $B'$ and $f'$ should be eliminated from (\ref{239}). Using (\ref{242}) and (\ref{247}) to eliminate $B'$ and $f'$ gives  
\begin{eqnarray}\label{250}
C'' 
&=& \frac{C'^2}{C}\Big(1+\frac{2\omega^2}{r^2}\Big) -\frac{2\omega\omega'C'}{r^2} -\frac{C'}{r} +\frac{C\omega'^2}{2r^2}\nonumber\\
&& -\frac{\varepsilon B^2\sin^2f}{\lambda^2}\Big[\frac{n^2}{C^2}\Big(1+\frac{\omega^2}{r^2}\Big) +\frac{2nmk\omega}{r^2} 
+\frac{C^2m^2k^2}{r^2}\Big]\nonumber\\
&&\times \Big(\frac{C}{2} -rC'\Big) \nonumber\\
&& +\frac{\varepsilon \sin^2f}{2}\Big[\frac{n^2}{C^2}\Big(\frac{\omega^2}{r^2}-1\Big)+\frac{2nmk\omega}{r^2} +\frac{C^2m^2k^2}{r^2}\Big]\nonumber\\
&&\Big(\frac{CB^2}{\lambda^2}-4K_s \left\{\frac{A'C}{rA} +\frac{\omega\omega'C'}{r^2} +\frac{C'}{r} -\frac{C'^2}{C}\Big(1+\frac{\omega^2}{r^2}\Big) \right.\nonumber\\
&&\left.-\frac{\omega'^2C}{4r^2} +\frac{\varepsilon B^2C\sin^2f}{2\lambda^2}\Big[\frac{n^2}{C^2}\Big(1+\frac{\omega^2}{r^2}\Big) \right.\nonumber\\
&&\left.+\frac{2nmk\omega}{r^2} 
+\frac{C^2m^2k^2}{r^2}\Big]\right\}\left\{\frac{\varepsilon}{2\lambda^2} -2\varepsilon K_s\sin^2f\right.\nonumber\\
&&\left.\times\Big[\frac{n^2}{C^2}\Big(1+\frac{\omega^2}{r^2}\Big) 
+\frac{2nmk\omega}{r^2} +\frac{C^2m^2k^2}{r^2}\Big]\right\}^{-1}\Big)
\end{eqnarray}

Similarly, $B'$ and $f'$ need to be eliminated from (\ref{238}). Using (\ref{242}) and (\ref{247}) to eliminate them gives 
\begin{eqnarray}\label{251}
\omega'' 
&=& \frac{\omega'}{r} + \frac{4\omega C'^2}{C^2}\Big(1+\frac{\omega^2}{r^2}\Big)-\frac{4\omega^2\omega'C'}{r^2C} - \frac{4\omega C'}{rC} +\frac{\omega\omega'^2}{r^2}\nonumber\\
&& - \frac{\varepsilon B^2(\omega - r\omega')\sin^2f}{\lambda^2}\Big[\frac{n^2}{C^2}\Big(1+\frac{\omega^2}{r^2}\Big) + \frac{2nmk\omega}{r^2} \nonumber\\
&& + \frac{C^2m^2k^2}{r^2}\Big] +\varepsilon \sin^2f\Big[\Big(\frac{n^2}{C^2} +\frac{2nmk}{\omega}\Big)\Big(1+\frac{\omega^2}{r^2}\Big) \nonumber\\
&&+\frac{C^2m^2k^2}{r^2}\Big]\Big(\frac{\omega B^2}{\lambda^2} - 4K_s \Big\{\frac{\omega A'}{rA} + \frac{\omega^2\omega'C'}{r^2C}\nonumber\\
&& + \frac{\omega C'}{rC} - \frac{\omega C'^2}{C^2}\Big(1+\frac{\omega^2}{r^2}\Big) - \frac{\omega \omega'^2}{4r^2}\nonumber\\
&& + \frac{\varepsilon \omega B^2\sin^2f}{2\lambda^2}\Big[\frac{n^2}{C^2}\Big(1+\frac{\omega^2}{r^2}\Big) + \frac{2nmk\omega}{r^2} + \frac{C^2m^2k^2}{r^2}\Big]\Big\}\nonumber\\
&&\times \Big\{\frac{\varepsilon}{2\lambda^2} - 2\varepsilon K_s\sin^2f\Big[\frac{n^2}{C^2}\Big(1+\frac{\omega^2}{r^2}\Big) + \frac{2nmk\omega}{r^2} \nonumber\\
&&+\frac{C^2m^2k^2}{r^2}\Big]\Big\}^{-1}\Big)
\end{eqnarray}

The system of equations to be solved is therefore
\begin{eqnarray}\label{252}
B'
&=& \frac{BA'}{A} + \frac{r\varepsilon B^3}{\lambda^2}\Big[\frac{n^2}{C^2}\Big(1+\frac{\omega^2}{r^2}\Big)  + \frac{2nmk\omega}{r^2} \nonumber\\
&&+\frac{C^2m^2k^2}{r^2}\Big]\sin^2f
\end{eqnarray}
\begin{eqnarray}\label{253}
f'
&=& \pm \Big\{\frac{\varepsilon}{2\lambda^2} - 2\varepsilon K_s\Big[\frac{n^2}{C^2}\Big(1+\frac{\omega^2}{r^2}\Big) + \frac{2nmk\omega}{r^2} \nonumber\\
&&+\frac{C^2m^2k^2}{r^2}\Big]\sin^2f\Big\}^{-1/2}\Big\{\frac{A'}{rA}+ \frac{\omega\omega'C'}{r^2C} + \frac{C'}{rC} \nonumber\\
&&-\frac{C'^2}{C^2}\Big(1+\frac{\omega^2}{r^2}\Big)  - \frac{\omega'^2}{4r^2} +\frac{\varepsilon B^2}{2\lambda^2}\Big[\frac{n^2}{C^2}\Big(1+\frac{\omega^2}{r^2}\Big) \nonumber\\
&&+\frac{2nmk\omega}{r^2} + \frac{C^2m^2k^2}{r^2}\Big]\sin^2f\Big\}^{1/2}
\end{eqnarray}
\begin{eqnarray}\label{254}
A'' 
&=& -\frac{A'}{r} + \frac{A'^2}{A} +\varepsilon\sin^2f\Big[\frac{n^2}{C^2}\Big(1+\frac{\omega^2}{r^2}\Big) + 
\frac{2nmk\omega}{r^2} \nonumber\\
&&+\frac{C^2m^2k^2}{r^2}\Big]\Big(\frac{rA'B^2}{\lambda^2} - 2K_s\Big\{\frac{A'}{r}+ \frac{\omega A\omega'C'}{r^2C} + \frac{AC'}{rC} \nonumber\\
&& -\frac{AC'^2}{C^2}\Big(1+\frac{\omega^2}{r^2}\Big)  - \frac{A\omega'^2}{4r^2} + \frac{\varepsilon AB^2}{2\lambda^2}\Big[\frac{n^2}{C^2}\Big(1+\frac{\omega^2}{r^2}\Big) \nonumber\\
&& +\frac{2nmk\omega}{r^2} + \frac{C^2m^2k^2}{r^2}\Big]\sin^2f\Big\} \nonumber\\
&&\times\Big\{\frac{\varepsilon}{2\lambda^2} - 2\varepsilon K_s\Big[\frac{n^2}{C^2}\Big(1+\frac{\omega^2}{r^2}\Big) + \frac{2nmk\omega}{r^2} + \frac{C^2m^2k^2}{r^2}\Big]\nonumber\\
&&\sin^2f\Big\}^{-1}\Big)
\end{eqnarray}
\begin{eqnarray}\label{255}
C'' 
&=& \frac{C'^2}{C}\Big(1+\frac{2\omega^2}{r^2}\Big) - \frac{2\omega\omega'C'}{r^2} - \frac{C'}{r} + \frac{C\omega'^2}{2r^2}\nonumber\\
&& - \frac{\varepsilon B^2\sin^2f}{\lambda^2}\Big[\frac{n^2}{C^2}\Big(1+\frac{\omega^2}{r^2}\Big) + \frac{2nmk\omega}{r^2} + \frac{C^2m^2k^2}{r^2}\Big]\nonumber\\
&&\times \Big(\frac{C}{2} - rC'\Big)  + \frac{\varepsilon \sin^2f}{2}\Big[\frac{n^2}{C^2}\Big(\frac{\omega^2}{r^2}-1\Big)+\frac{2nmk\omega}{r^2} \nonumber\\
&&+\frac{C^2m^2k^2}{r^2}\Big]\Big(\frac{CB^2}{\lambda^2}-4K_s \Big\{\frac{A'C}{rA} + \frac{\omega\omega'C'}{r^2}  +\frac{C'}{r} \nonumber\\
&& -\frac{C'^2}{C}\Big(1+\frac{\omega^2}{r^2}\Big) - \frac{\omega'^2C}{4r^2}  + \frac{\varepsilon B^2C\sin^2f}{2\lambda^2}\Big[\frac{n^2}{C^2}\Big(1+\frac{\omega^2}{r^2}\Big) \nonumber\\
&& +\frac{2nmk\omega}{r^2} + \frac{C^2m^2k^2}{r^2}\Big]\Big\}\Big\{\frac{\varepsilon}{2\lambda^2} - 2\varepsilon K_s \sin^2f\nonumber\\
&&\times \Big[\frac{n^2}{C^2}\Big(1+\frac{\omega^2}{r^2}\Big) + \frac{2nmk\omega}{r^2} + \frac{C^2m^2k^2}{r^2}\Big]\Big\}^{-1}\Big)
\end{eqnarray}
\begin{eqnarray}\label{256}
\omega'' 
&=& \frac{\omega'}{r} + \frac{4\omega C'^2}{C^2}\Big(1+\frac{\omega^2}{r^2}\Big)-\frac{4\omega^2\omega'C'}{r^2C} - \frac{4\omega C'}{rC} +\frac{\omega\omega'^2}{r^2}\nonumber\\
&& - \frac{\varepsilon B^2(\omega - r\omega')\sin^2f}{\lambda^2}\Big[\frac{n^2}{C^2}\Big(1+\frac{\omega^2}{r^2}\Big) + \frac{2nmk\omega}{r^2} \nonumber\\
&& +\frac{C^2m^2k^2}{r^2}\Big]  +\varepsilon \sin^2f \Big[\Big(\frac{n^2}{C^2} +  \frac{2nmk}{\omega}\Big)\Big(1+\frac{\omega^2}{r^2}\Big) \nonumber\\
&&+\frac{C^2m^2k^2}{r^2}\Big] \Big(\frac{\omega B^2}{\lambda^2} - 4K_s \Big\{\frac{\omega A'}{rA} + \frac{\omega^2\omega'C'}{r^2C} +\frac{\omega C'}{rC} \nonumber\\
&& -\frac{\omega C'^2}{C^2}\Big(1+\frac{\omega^2}{r^2}\Big) - \frac{\omega \omega'^2}{4r^2} 
+\frac{\varepsilon \omega B^2\sin^2f}{2\lambda^2}\nonumber\\
&&\times \Big[\frac{n^2}{C^2}\Big(1+\frac{\omega^2}{r^2}\Big) + \frac{2nmk\omega}{r^2} + \frac{C^2m^2k^2}{r^2}\Big]\Big\}\nonumber\\
&&\times \Big\{\frac{\varepsilon}{2\lambda^2} - 2\varepsilon K_s\sin^2f\Big[\frac{n^2}{C^2}\Big(1+\frac{\omega^2}{r^2}\Big) + \frac{2nmk\omega}{r^2} \nonumber\\
&&+\frac{C^2m^2k^2}{r^2}\Big]\Big\}^{-1}\Big)
\end{eqnarray}

The plus/minus sign $\pm$ in eq.(\ref{253}) indicates that there is a choice of signs for $f'$. But since the boundary conditions on $f$ will require that $f(0)=\pi$ and $\lim_{r\rightarrow\infty}f(r)=0$, it is likely that $f'<0$ for all values of $r$. 

\section{Solutions of the Einstein field equations for small values of $r$ ($r\rightarrow 0$)}
In order to integrate the Einstein equations numerically, it is necessary to make some assumptions about the behaviour of the metric functions $A,~B,~C$ and $\omega$ and the field function $f$ for small values of $r$. 

The Einstein field equations (\ref{252}) to (\ref{256}) place constraints on the values of the constants appearing in the boundary conditions, both for small values of $r$, i.e. for $r\rightarrow 0$, and for large values of $r$, i.e. for $r\rightarrow\infty$. The Einstein field equations for small values of $r$ can be generated by substituting (\ref{252})-(\ref{256}) into (\ref{242})-(\ref{251}).
Note that the field equation for $\omega$ contains no extra information about the expansion coefficients for small values of $r$. We set the coefficient $C_0$ equal to $B_0$, as the condition $C_0=B_0$ is necessary to avoid a conical singularity at $r=0$.  

Remembering that the solution for the Skyrmion field, $f$, in the absence of self-gravity or a twist is known to be
\begin{equation}\label{354}
f(r) = \pi - 2K^{1/2}r^n + \frac{2}{3}K^{3/2}r^{3n} + ...
\end{equation}
and that the Skyrmion field was assumed to have the small-$r$ expansion
\begin{equation}\label{355}
f(r) = \pi + ar + cr^3 + ...
\end{equation}

It is evident that (\ref{354}) and (\ref{355}) are consistent only if the field is in its ground state, that is, has $n=1$. So, the relevant condition for $n^2B_0^2-C_0^2=0$, when we set $B_0=C_0$ to avoid a conical singularity at $r=0$, is 
\begin{equation}\label{356}
1-n^2=0\rightarrow n = 1
\end{equation}
So, the condition $n=1$ is the one that we will choose from (\ref{356}). The other possible choice, $n=-1$, is equivalent to $n=1$ under a reversal $\theta\rightarrow -\theta$ of the sign of the angular coordinate $\theta$.

Using relations $n=1$ and $B_0=C_0$, we obtain 
\begin{equation}\label{365}
C_2 = -\frac{\varepsilon K_s a^4}{2C_0}
\end{equation}
\begin{equation}\label{366}
A_2 = \frac{\varepsilon K_s a^4 A_0}{2C_0^2}
\end{equation}
\begin{eqnarray}\label{367}
B_2 
&=& \frac{\varepsilon a^2 B_0}{2}\Big(\frac{1}{\lambda^2} +\frac{K_s a^2}{B_0^2}\Big)
\end{eqnarray}
Substituting (\ref{366}) into $A(r)$, we obtain
\begin{equation}\label{368}
A(r) = A_0 +\frac{\varepsilon K_s a^4 A_0}{2C_0^2}r^2 + ...
\end{equation}
Substituting (\ref{367}) into $B(r)$, we obtain
\begin{eqnarray}\label{369}
B(r) 
&=& B_0 + \frac{\varepsilon a^2 B_0}{2}\Big(\frac{1}{\lambda^2} +\frac{K_s a^2}{B_0^2}\Big) r^2 + ...
\end{eqnarray}
Substituting (\ref{365}) into $C(r)$ and using $B_0=C_0$, we obtain
\begin{eqnarray}\label{370}
C(r) 
&=& B_0r -\frac{\varepsilon K_s a^4}{2B_0}r^3 + ...
\end{eqnarray}
The expansion $f(r)$ still reads 
\begin{equation}\label{371}
f(r) = \pi + ar + ...
\end{equation}
and the expansion $\omega(r)$ as
\begin{equation}\label{372}
\omega(r) = \omega_1r^2 + ...
\end{equation}

The constants $A_0$, $B_0$, $a$ and $\omega_1$ appearing in the expansions (\ref{368})-(\ref{372}) will be determined numerically, by matching the small-distance expansions with solutions that are asymptotically flat in the limit $r\rightarrow \infty$.

\section{Solutions of the Einstein field equations for large value of $r$ ($r\rightarrow\infty$)}
It is expected that at large values of $r$, the space-time surrounding the Skyrmion string will be locally flat (although possibly with an angle deficit $\Delta$ at infinity caused by the presence of the string along the axis of symmetry, and possibly also with a twist in the $\theta-z$ direction if the twist coefficient $mk$ is non-zero). 

This mean that the metric functions should satisfy
\begin{eqnarray}\label{373}
\lim_{r\rightarrow\infty}A(r) = 1,~~~
\lim_{r\rightarrow\infty}B(r) = 1
\end{eqnarray}
\begin{eqnarray}
\lim_{r\rightarrow\infty}\frac{C(r)}{r} = 1 -\frac{\Delta}{2\pi}=k_1
~~
\lim_{r\rightarrow\infty}\frac{\omega(r)}{r} = k_2
\end{eqnarray}

If we ignore $k_2$ for the moment, the asymptotic line element when $\Delta>0$ has the form
\begin{eqnarray}\label{377}
ds^2=dt^2-dr^2-\left(1-\frac{\Delta}{2\pi}\right)^2r^2d\theta^2 -\left(1-\frac{\Delta}{2\pi}\right)^{-2}dz^2 \nonumber\\
\end{eqnarray}
Here, $\Delta$ can be eliminated from the $dz$ term by defining a new vertical coordinates $Z=[1-\Delta/(2\pi)]^{-1}z$, so that
\begin{eqnarray}\label{378}
ds^2=dt^2-dr^2-\left(1-\frac{\Delta}{2\pi}\right)^2r^2d\theta^2-dZ^2.
\end{eqnarray}
But, $\Delta$ cannot be eliminated from the $d\theta$ term without destroying the standard $2\pi$-periodicity of the angular coordinate $\theta$. 

In fact, the circumference $C(r)$ of a horizontal circle of radius $r$ in this space-time is calculated by letting $\theta$ run from 0 to $2\pi$, so that
\begin{eqnarray}\label{379}
C(r) &=& 2\pi\left(1-\frac{\Delta}{2\pi}\right)r=(2\pi-\Delta)r
\end{eqnarray}
In effect, therefore, an angle of size $\Delta$ is missing from all horizontal circles in the limit of large $r$, which is why $\Delta$ is usually called the "angle deficit" of the space-time.

If $k_2$ is also non-zero, then the asymptotic line element has the even more complicated form ($\frac{\omega}{r}\rightarrow 0$ for $r\rightarrow\infty$)
\begin{equation}\label{380}
ds^2=dt^2-dr^2-k_1^2r^2d\theta^2+2k_2r~d\theta~dz -(1+k_2^2)k_1^{-2}dz^2
\end{equation}
In terms of $Z=k_1^{-1}z$, this (\ref{380}) can be written as ($\frac{\omega}{r}\rightarrow 0$ for $r\rightarrow\infty$)
\begin{eqnarray}\label{381}
ds^2=dt^2-dr^2-(k_1r~d\theta-k_2~dZ)^2-dZ^2
\end{eqnarray}

Here, the circumferences of horizontal circles (which have $dZ=0$) are still shorter than $2\pi r$ by an amount $\Delta r$, but if $dZ\neq 0$ the lines of shortest distance between two points in the $\theta-Z$ plane do not have $d\theta=0$, but instead have $r~d\theta=k_1^{-1}k_2~dZ$ and so necessarily twist around the surface of a cylinder.

It is also expected that the deviations of the metric functions from their asymptotic values will be proporsional, to leading order, to the gravitational coupling constants $\varepsilon/\lambda^2$ and $\varepsilon K_s$ that, in eqs.(\ref{368}) and (\ref{369})-(\ref{372}), link the derivatives of $A,~B,~C$ and $\omega$ to the Skyrmion field function $f$. 

So, the metric functions will be expanded in the form
\begin{eqnarray*}
A(r) = 1 +\frac{\varepsilon}{\lambda^2}\alpha(r),~~
B(r) = 1 +\frac{\varepsilon}{\lambda^2}\beta(r)
\end{eqnarray*}
\begin{eqnarray}\label{382}
C(r) = k_1r\left[1+\frac{\varepsilon}{\lambda^2}\gamma(r)\right],~~
\omega(r) = \frac{\varepsilon}{\lambda^2}w(r)
\end{eqnarray}
where we expect that
\begin{eqnarray}\label{386}
\lim_{r\rightarrow\infty}\alpha(r)=\lim_{r\rightarrow\infty}\beta(r)=\lim_{r\rightarrow\infty}\gamma(r)=0
\end{eqnarray}
and that
\begin{eqnarray}\label{387}
\frac{\varepsilon}{\lambda^2}\lim_{r\rightarrow\infty} \frac{w(r)}{r} = k_2
\end{eqnarray}
although there is no guarantee that $k_2$ will be non-zero.

On writing $K_s=\hat{K}/\lambda^2$ and expanding all five field equations (\ref{368})-(\ref{372}) to leading order in $\varepsilon/\lambda^2$ we get
\begin{eqnarray}\label{388}
\beta'=\alpha'+\frac{1}{r}\left(\frac{n^2}{k_1^2}+k_1^2m^2k^2r^2\right)\sin^2f
\end{eqnarray}
\begin{eqnarray}\label{389}
f'
&=&\pm\left\{\frac{1}{2} +\frac{2\hat{K}}{r^2}\left(\frac{n^2}{k_1^2} +k_1^2m^2k^2r^2\right)\sin^2f\right\}^{-1/2}\nonumber\\
&&\times\left\{\frac{1}{r}(\alpha'-\gamma') +\frac{1}{2r^2}\left(\frac{n^2}{k_1^2} +k_1^2m^2k^2r^2\right)\sin^2f\right\}^{1/2}\nonumber\\
\end{eqnarray}
\begin{eqnarray}\label{390}
\alpha''
&=& -\frac{\alpha'}{r} +\frac{2\hat{K}}{r^2}\left(\frac{n^2}{k_1^2} +k_1^2m^2k^2r^2\right)\sin^2f\nonumber\\
&&\times\left[\frac{\alpha'-\gamma'}{r} +\frac{1}{2r^2}\left(\frac{n^2}{k_1^2} +k_1^2m^2k^2r^2\right)\sin^2f\right]\nonumber\\
&&\times\left[\frac{1}{2} +\frac{2\hat{K}}{r^2}\left(\frac{n^2}{k_1^2} +k_1^2m^2k^2r^2\right)\sin^2f\right]^{-1}
\end{eqnarray}
\begin{eqnarray}\label{391}
\gamma''
&=& -\frac{\gamma'}{r}  +k_1^2m^2k^2\sin^2f \nonumber\\
&& +\frac{2\hat{K}}{r^2}\left(-\frac{n^2}{k_1^2} +k_1^2m^2k^2r^2\right)\sin^2f\nonumber\\
&&\times\left\{\frac{\alpha'-\gamma'}{r} +\frac{1}{2r^2}\left(\frac{n^2}{k_1^2} +k_1^2m^2k^2r^2\right)\sin^2f\right\}\nonumber\\
&&\times\left[\frac{1}{2} +\frac{2\hat{K}}{r^2}\left(\frac{n^2}{k_1^2} +k_1^2m^2k^2r^2\right)\sin^2f\right]^{-1}
\end{eqnarray}
and
\begin{eqnarray}\label{392}
w''
&=& \frac{w'}{r} +2nmk~\sin^2f\nonumber\\
&&\times\left[1 +4\hat{K}\left\{\frac{\alpha'-\gamma'}{r} \right.\right.\nonumber\\
&&\left.\left.+\frac{1}{2r^2}\left(\frac{n^2}{k_1^2} +k_1^2m^2k^2r^2\right)\sin^2f\right\}\right.\nonumber\\
&&\left.\times\left\{\frac{1}{2} +\frac{2\hat{K}}{r^2}\left(\frac{n^2}{k_1^2} +k_1^2m^2k^2r^2\right)\sin^2f\right\}^{-1}\right]\nonumber\\
\end{eqnarray}

In the asymptotic region, where $f$ is small and self-gravitating forces are correspondingly weak, another expectation is that $f$ will behave as it does at large distances from the axis in the non-gravitating cases (although possibly with minor variations due to the presence of the angle deficit and the twist). 

Recall from previous chapter that $f$ goes to zero like $r^{-n}$ at large distances if there is no twist - or at least it would if any solutions regular at both $r=0$ and $r=\infty$ existed at all. And it was shown that solutions with a twist go to zero like $r^{-1/2}\exp(-mkr)$.

\section{Metric and field functions for large value of $r$: non-twisting case ($\omega = 0$)}
If $mk=0$ then eq.(\ref{392}) and also the full field equation for $\omega$ are trivially satisfied if $\omega=0$ or equivalently $w=0$ everywhere. 

In summary, in the twist-free case, the metric functions and the Skyrmion field function $f$ have the following asymptotic expansions for large values of $r$
\begin{equation}\label{406}
A(r)\approx 1 +\frac{1}{2} \frac{\varepsilon K_s}{(2n/k_1+1)^{2}} \left(\frac{n^2}{k_1^2}\right)^2 \frac{f_0^4}{r^{4n/k_1+2}}
\end{equation}
\begin{equation}\label{407}
B(r)\approx 1 -\frac{\varepsilon}{2\lambda^2} \frac{n}{k_1} \frac{f_0^2}{r^{2n/k_1}}
\end{equation}
\begin{equation}\label{408}
C(r)\approx k_1r\left[1 -\frac{\varepsilon K_s}{2(2n/k_1+1)^{2}} \left(\frac{n^2}{k_1^2}\right)^2 \frac{f_0^4}{r^{4n/k_1+2}}\right]
\end{equation}
and
\begin{equation}\label{409}
f(r)\approx \frac{f_0}{r^{n/k_1}}
\end{equation}

It might seem surprising that $f$ goes to zero like $r^{-n/k_1}$ if self-gravity is included, whereas it goes to zero like $r^{-n}$ in the absence of self gravity. The difference is of course due to the angle deficit - if there is one, which can only be established once the field equations have been integrated numerically. 

In fact, it turns out that powers of $r^{1/k_1}$ appear naturally in a conical space-time with a space-like sector of the form
\begin{eqnarray}\label{410}
dl^2=dr^2 +k_1^2r^2d\theta^2
\end{eqnarray}
which is what we expect the asymptotic line element to contain in the twist-free case, as we have seen.

The line element $dl^2$ can be rewritten in conformally-flat form by defining a new pair of coordinates $X$ and $Y$ via the equations
\begin{eqnarray}\label{411}
r\cos\theta=R^{k_1-1}X~~~\text{and}~~~r\sin\theta=R^{k_1-1}Y
\end{eqnarray}
where $R=(X^2+Y^2)^{1/2}$. Since $r=R^{k_1}$ and $\theta=\tan^{-1}(Y/X)$ it follows immediately that
\begin{eqnarray*}
dr=k_1~R^{k_1-1}(X~dX+Y~dY) 
\end{eqnarray*}
and
\begin{eqnarray}\label{412}
r~d\theta=R^{k_1-1}(X~dY-Y~dX)
\end{eqnarray}
and so
\begin{eqnarray}\label{413}
dl^2=k_1^2~R^{2k_1}(dX^2+dY^2)
\end{eqnarray}

The line element $dl^2$ in this form is conformally flat, and the horizontal radius in the conformal coordinates $X$ and $Y$ is $R=r^{1/k_1}$. So, it can be seen from eq.(\ref{409}) that the leading term in the asymptotic expansion for $f$ is proportional to $R^{-n}$, rather than $r^{-n}$.

\section{Metric and field functions for large value of $r$: twisting case ($\omega\neq 0$)}
In summary, in the twisting case $(mk\neq 0$) the metric functions and the Skyrmion field function $f$ have the following asymptotic expansions for large values of $r$

\begin{equation}\label{437}
A(r)\approx 1 +\frac{1}{8}\varepsilon K_s~k_1^2m^2k^2 \frac{f_0^4}{r^{2}~e^{4k_1mkr}}
\end{equation}
\begin{equation}\label{438}
B(r)\approx 1-\frac{\varepsilon}{2\lambda^2}k_1mk \frac{f_0^2}{e^{2k_1mkr}}
\end{equation}
\begin{equation}\label{439}
C(r)\approx k_1r\left[1+\frac{\varepsilon}{4\lambda^2} \frac{f_0^2}{re^{2k_1mkr}}\right]
\end{equation}
\begin{equation}\label{440}
\omega(r)\approx \frac{\varepsilon}{2\lambda^2}\frac{n}{k_1^2mk} \frac{f_0^2}{r e^{2k_1mkr}}
\end{equation}
and
\begin{equation}\label{441}
f(r)\approx \frac{f_0}{\sqrt{r}~ e^{k_1mkr}}
\end{equation}

The asymptotic behaviour here of the twist function $\omega$ is potentially problematic, as $\omega$ is proportional to $(k_1mkr)^{-1}\exp(-2k_1mkr)$ and therefore diverges, at any fixed value of $r$, in the twist-free limit $mk\rightarrow 0$. This is simply a reflection of the fact that the asymptotic expansions used above in the twisting case are valid only for values $r>>1/(k_1mk)$.

\section{Proof that no non-twisting or twisting solutions extending from $r=0$ to $r=\infty$ exist}
The asymptotic expansions developed in the previous sections would apply to any self-gravitating vortex solutions that extend smoothly from $r=0$ to any asymptotically flat regime at $r=\infty$. However, it turns out that no such solutions actually exist - apart from the classical Linet cosmic strings, which have no twist and no non-zero stress-energy components except $T_t^t$ and $T_z^z$, with $T_t^t\equiv T_z^z$.

In retrospect, the failure of the baby Skyrmion vortex strings to admit consistent self-gravitating solutions is not very surprising, as they are infinitely long and so have an infinite energy. It is very unlikely then that they could be embedded in an asymptotically flat space-time. The Linet cosmic strings are an exception because their stress-energy tensor $T_{\nu}^{\mu}$ has a high degree of symmetry. 

In fact, $T_{\nu}^{\mu}$ is invariant under Lorentz boosts parallel to the $z$-azis, and it is simply shown that a radial gravitational force would break this symmetry. So, the Linet strings exert no gravitational force at all, and can be embedded in asymptotically flat space-time - although always with an angle deficit.

Now define a new dependent variable $U$ by
\begin{eqnarray}\label{447}
U\equiv \frac{rA'}{A}
\end{eqnarray}
Then
\begin{eqnarray}\label{448}
U'
&=& \left(\frac{rA'}{A}\right)' =\frac{rA''}{A} +\frac{A'}{A} -\frac{rA'^2}{A^2}
\end{eqnarray}
and the equation for $A''/A$ becomes
\begin{eqnarray}\label{449}
U'
&=& r\sin^2f\left[\frac{n^2}{C^2}\left(1+\frac{\omega^2}{r^2}\right) +\frac{2mnk\omega}{r^2} +\frac{C^2m^2k^2}{r^2}\right]\nonumber\\
&&\times \left(\frac{\varepsilon}{\lambda^2}UB^2 -2\varepsilon K_sf'^2\right)
\end{eqnarray}
Note here that
\begin{eqnarray}\label{450}
&&\frac{n^2}{C^2} +\frac{(n\omega +C^2mk)^2}{C^2r^2} \nonumber\\
&=& \frac{n^2}{C^2}\left(1+\frac{\omega^2}{r^2}\right) +\frac{2mnk\omega}{r^2} +\frac{C^2m^2k^2}{r^2}
\end{eqnarray}
is a strictly positive function.

Now, the space-time line element in the model is
\begin{eqnarray}\label{451}
ds^2
&=& A^2dt^2 -B^2dr^2 -C^2d\theta^2 +2\omega~d\theta~dz \nonumber\\
&&-\left(\frac{r^2+\omega^2}{C^2}\right)dz^2 
\end{eqnarray}
So $A$ is the scale factor for the timelike coordinate $t$, and must tend to finite non-zero values as $r\rightarrow 0$ and as $r\rightarrow \infty$, since otherwise the space-time metric will be singular. This in turn means that $U$ must tend to 0 as $r\rightarrow 0$ and as $r\rightarrow \infty$. 

For suppose that $U\rightarrow u_0\neq 0$ as $r\rightarrow 0$. Then for small values of $r$,
\begin{eqnarray}\label{452}
\frac{A'}{A}
&=& (\ln A)'\approx \frac{u_0}{r}
\end{eqnarray}
and so
\begin{eqnarray}\label{453}
\ln A\approx u_0\ln r +\ln A_0
\end{eqnarray}
(where $A_0$ is an integration constant), or equivalently
\begin{eqnarray}\label{454}
A\approx A_0r^{u_0}
\end{eqnarray}
That is, since $u_0\neq 0$, $A$ must either diverge (if $u_0<0$) or go to 0 (if $u_0>0$) as $r\rightarrow 0$. 

In both cases, the metric will be singular at $r=0$, so we conclude that $U\rightarrow 0$ as $r\rightarrow 0$. Similarly, if we suppose that $U\rightarrow u_\infty\neq 0$ as $r\rightarrow\infty$ then for large values of $r$
\begin{eqnarray}\label{455}
A\approx A_\infty r^{u_\infty}
\end{eqnarray}
for some integration constant $A_\infty$. $A$ must either diverge (if $u_\infty>0$) or go to 0 (if $u_\infty <0$) as $r\rightarrow\infty$. So, we conclude that $U\rightarrow 0$ as $r\rightarrow\infty$.

Consider now the behaviour of the equation
\begin{eqnarray}\label{456}
U'
&=& r\sin^2f\left[\frac{n^2}{C^2}\left(1+\frac{\omega^2}{r^2}\right) +\frac{2mnk\omega}{r^2} +\frac{C^2m^2k^2}{r^2}\right]\nonumber\\
&&\times \left(\frac{\varepsilon}{\lambda^2}UB^2 +2\varepsilon K_sf'^2 \right)
\end{eqnarray}
in the limit of small $r$. We know that $f\rightarrow\pi$, $B\rightarrow B_0>0$, $C\approx B_or$ and $\omega/r\rightarrow 0$ as $r\rightarrow 0$. So
\begin{eqnarray}\label{457}
U'
&\approx & \frac{(f-\pi)^2}{r}\frac{n^2}{B_0^2}\left(\frac{\varepsilon}{\lambda^2}UB_0^2 +2\varepsilon K_s f'^2\right)
\end{eqnarray}
for small values of $r$. Futhermore, by l'Hopital's rule
\begin{eqnarray}\label{458}
\frac{f-\pi}{r}\approx f'
\end{eqnarray}
and so (after dividing through by $U$)
\begin{eqnarray}\label{459}
\frac{U'}{U}\approx rf'^2\frac{n^2}{B_0^2}\left(\frac{\varepsilon}{\lambda^2}B_0^2 +2\varepsilon K_s\frac{f'^2}{U}\right)
\end{eqnarray}

Here $U'/U=(\ln|U|)'$, and so if $U\rightarrow 0$ as $r\rightarrow 0$, we must have $U'/U\rightarrow\infty$ (as $\ln|U|\rightarrow-\infty$ and so its derivative must tend to $\infty$ as $r\rightarrow 0$ if $\ln |U|$ is to be finite for $r>0$). But if $f'$ remains bounded as $r\rightarrow 0$ the only possible divergent term on the right of this equation is $f'^2/U$, whereas if $f'$ diverges as $r\rightarrow 0$ then clearly $|2\varepsilon K_s f'^2/U|>> \varepsilon B_0^2/\lambda^2$. In both cases then
\begin{eqnarray}\label{460}
\frac{U'}{U}\approx rf'^2\frac{n^2}{B_0^2}\left(2\varepsilon K_s\frac{f'^2}{U}\right)
\end{eqnarray}
and since $U'/U\rightarrow\infty$, $U$ must have the same sign as $K_s$ for small non-zero values of $r$. But it is clear that from this equation that $U'$ also must have the same sign as $K_s$ for small non-zero values of $r$.

Returning now to the full equation for $U'$, namely
\begin{eqnarray}\label{461}
U'
&=& r\sin^2f\left[\frac{n^2}{C^2}\left(1+\frac{\omega^2}{r^2}\right) +\frac{2mnk\omega}{r^2} +\frac{C^2m^2k^2}{r^2}\right]\nonumber\\
&&\times \left(\frac{\varepsilon}{\lambda^2}UB^2 +2\varepsilon K_sf'^2\right)
\end{eqnarray}
It is evident that the right-hand side consists of a strictly positive function times the sum of two terms, $\varepsilon UB^2/\lambda^2$ and $2\varepsilon K_sf'^2$, which both have the same sign as $K_s$ for small non-zero values of $r$. 

It follows therefore that $U'$ can never change sign [from sgn($K_s$) to sgn($-K_s$)], because this could only happen if $U$ were negative, and $U$ cannot change sign before $U'$ does. So, $U'$ and $U$ will continue to be positive for all $r>0$, and it is not possible for $U$ to go back to 0 as $r\rightarrow\infty$.

\section{Non-twisting and twisting solutions of the Einstein field equations with a finite radius}
Given that we know now that there are no non-trivial solutions to the baby Skyrmion field equations - with or without a twist - that extend continuously from $r=0$ to $r=\infty$. Then the next task is to search for solutions with a boundary at a finite radius $r=r_b$. What this means is that the metric functions $A$, $B$, $C$ and $\omega$ and the Skyrmion field $f$ satisfy the Einstein equations on the domain $0\leq r<r_b$. But the field energy is zero, and so the metric satisfies the vacuum Einstein equations, on the domain $r_b<r<\infty$.

As previously, the line element on the domain $0\leq r<r_b$ is assumed to have the standard form
\begin{eqnarray}\label{462}
ds_{\text{in}}^2
&=& A^2dt^2 -B^2dr^2 -C^2d\theta^2 +2\omega~d\theta~dz \nonumber\\
&& -\left(\frac{r^2+\omega^2}{C^2}\right)dz^2 
\end{eqnarray}

The line element in the vacuum region $r_b< r<\infty$ can be generated by setting $f\equiv 0$ in the field equations. This gives: (i)
\begin{eqnarray}\label{463}
B'=\frac{BA'}{A}~~~\text{or equivalently}~~~(\ln B)'=(\ln A)'
\end{eqnarray}
and so $A/B$ is constant: (ii)
\begin{eqnarray}\label{464}
A''=-\frac{A'}{r} +\frac{A'^2}{A^2}~~~\text{or equivalently}~~[r(\ln A)']'=0
\end{eqnarray}
and so $A=A_*r^q$ for some choice of constants $A_*$ and $q$: (iii)
\begin{eqnarray}\label{465}
C''
&=& \frac{C'^2}{C}\left(1+\frac{2\omega^2}{r^2}\right) -\frac{2\omega\omega'C'}{r^2} -\frac{C'}{r} +\frac{C\omega'^2}{2r^2}
\end{eqnarray}
and (iv)
\begin{eqnarray}\label{466}
\omega''
&=& \frac{\omega'}{r} +\frac{4\omega C'^2}{C^2}\left(1+\frac{\omega^2}{r^2}\right) -\frac{4\omega^2\omega'C'}{r^2C}  -\frac{4\omega C'}{rC} \nonumber\\
&& +\frac{\omega\omega'^2}{r^2}
\end{eqnarray}
Also, the equation for $f'$ becomes
\begin{eqnarray}\label{467}
f'
&=& \pm\left(\frac{\varepsilon}{2\lambda^2}\right)^{-1/2}\left[\frac{A'}{rA} +\frac{\omega\omega'C'}{r^2C} +\frac{C'}{rC} \right.\nonumber\\
&&\left.-\frac{C'^2}{C^2}\left(1+\frac{\omega^2}{r^2}\right) -\frac{\omega'^2}{4r^2}\right]^{1/2}
\end{eqnarray}
and so (v)
\begin{eqnarray}\label{468}
0=\frac{A'}{rA} +\frac{\omega\omega'C'}{r^2C} +\frac{C'}{rC} -\frac{C'^2}{C^2}\left(1+\frac{\omega^2}{r^2}\right) -\frac{\omega'^2}{4r^2} \nonumber\\
\end{eqnarray}
as $f'\equiv 0$ in the vacuum region.

For the metric to remain non-singular as $r\rightarrow \infty$, we require $A$ to remain finite and non-zero. So, we must have $q=0$ and therefore $A=A_*$ and $B=B_*$ for some constants $A_*$ and $B_*$. Equation (v) then reads
\begin{eqnarray}\label{469}
\frac{C'^2}{C^2}\left(1+\frac{\omega^2}{r^2}\right) -\frac{\omega\omega'C'}{r^2C} -\frac{C'}{rC} +\frac{\omega'^2}{4r^2} = 0
\end{eqnarray}
which when substituted into (iv) gives
\begin{eqnarray}\label{470}
\omega''=\frac{\omega'}{r}~~~\text{or equivalently}~~~\left(\frac{\omega'}{r}\right)'=0
\end{eqnarray}
and so 
\begin{eqnarray}\label{471}
\omega &=&\omega_*+\Omega_*r^2
\end{eqnarray}
for some choice of constants $\omega_*$ and $\Omega_*$. 

Substituting this expression for $\omega$ back into (v) gives
\begin{eqnarray}\label{472}
0
&=&\frac{C'^2}{C^2}\left(1+2\omega_*\Omega_*+\Omega_*^2r^2+\frac{\omega_*^2}{r^2}\right) \nonumber\\
&&-(1+2\omega_*\Omega_*+2\Omega_*^2r^2)\frac{C'}{rC} +\Omega_*^2
\end{eqnarray}
This is a quadratic equation for $C'/C$, with solution
\begin{eqnarray}\label{473}
\frac{C'}{C}&=& \frac{1}{2}r\frac{1+2\omega_*\Omega_*+2\Omega_*^2r^2\pm\sqrt{1+4\omega_*\Omega_*}}{r^2+(\omega_*+\Omega_*r^2)^2}
\end{eqnarray}
If $\Omega_*\neq 0$ this equation integrates to give
\begin{eqnarray}\label{474}
C
&=& C_*\left(\frac{1+2\omega_*\Omega_* +2\Omega_*^2r^2 \mp\sqrt{1+4\omega_*\Omega_*}}{1+2\omega_*\Omega_* +2\Omega_*^2r^2 \pm\sqrt{1+4\omega_*\Omega_*}}\right)^{1/4}\nonumber\\
&&\times [r^2 +(\omega_*+\Omega_*r^2)^2 ]^{1/4}
\end{eqnarray}
where $C_*$ is a constant. 

In particular, in the limit as $r\rightarrow \infty$, the metric function $C$ has the asymptotic form
\begin{eqnarray}\label{475}
C\approx C_*|\Omega_*|^{1/2}r
\end{eqnarray}
and the metric function outside $dz^2$ in the vacuum line element has the form
\begin{eqnarray}\label{476}
\frac{r^2+\omega^2}{C^2}\approx C_*^{-2}|\Omega_*|r^2
\end{eqnarray}
which diverges as $r\rightarrow\infty$. 

In order for $z$ to remain as a vertical Cartesian coordinate, we need therefore to choose $\Omega_*=0$. Then $\omega =\omega_*$ and the equation for $C'/C$ reads
\begin{eqnarray}\label{477}
\frac{C'}{C}
&=& \frac{r}{r^2+\omega_*^2}
\end{eqnarray}
This equation solves to give
\begin{eqnarray}\label{478}
C=C_*(r^2+\omega_*^2)^{1/2}
\end{eqnarray}
which also means that
\begin{eqnarray}\label{479}
\frac{r^2+\omega^2}{C^2}=C_*^{-2}
\end{eqnarray}
[Note that a second possible root of (v) in the case $\Omega_*=0$ is $C'/C=0$, which is satisfied if $C$ is constant. But this solution should be rejected if $\theta$ is to remain an angular coordinate as $r\rightarrow\infty$, because $C$ should then be proportional to $r$.]

In summary, the line element in the vacuum region $r_b<r<\infty$ is
\begin{eqnarray}\label{480}
ds_{\text{out}}^2
&=&A_*^2dt^2 -B_*^2dr^2 -C_*^2(r^2+\omega_*^2)d\theta^2 +2\omega_*~d\theta~ dz \nonumber\\
&& -C_*^{-2}~dz^2
\end{eqnarray}
where $A_*$, $B_*$, $C_*$ and $\omega_*$ are all constants. This line element can be rewritten in the form
\begin{eqnarray}\label{481}
ds_{\text{out}}^2
&=& A_*^2~dt^2 -B_*^2~dr^2 -C_*^2r^2~d\theta^2 -(C_*^{-1}dz \nonumber\\
&&-C_*\omega_*~d\theta)^2
\end{eqnarray}
which indicates that the exterior vacuum space-time contains an angle deficit 
\begin{eqnarray}\label{482}
\Delta = 2\pi(1-C_*/B_*)
\end{eqnarray}
if $C_*<B_*$, and also a twist if $\omega_*\neq 0$.

In order for the two line elements $ds_{\text{in}}^2$ and $ds_{\text{out}}^2$ to together represent a viable baby Skyrmion string solution, they must be matched at the boundary radius $r=r_b$. The various methods that have historically been used to match two solutions of Einstein's equations across a non-null (that is, time-like or space-like) boundary surface were long ago listed and compared in a paper by Bonnor and Vickers \cite{bon}. 

There are three standard methods for doing this (the Darmois, O'Brien and Synge, and Lichnerowicz method), which involve imposing what are called junction conditions on the two line elements $ds_{\text{in}}^2$ and $ds_{\text{out}}^2$, but in many cases the three methods are equivalent. 

In the case we are considering they are in fact equivalent, and the junction conditions consist of two parts:
\begin{itemize}
\item[(a)] The two line elements on the boundary surface $r=r_b$ must be continuous across the surface, which means that
\begin{eqnarray}\label{483}
&& A_b^2dt^2 - C_b^2d\theta^2 +2\omega_b d\theta~dz -\frac{r_b^2+\omega_b^2}{C_b^2} dz^2 \nonumber\\
&=& A_*^2dt^2 -C_*^2(r_b^2+\omega_*^2)d\theta^2 +2\omega_*d\theta~dz \nonumber\\
&&-~C_*^{-2}dz^2
\end{eqnarray}
where the subscript $b$ in the left-hand expression indicates that the corresponding function is evaluated at $r=r_b$, so that $A_b\equiv A(r_b)$ etc. Hence, the first set of junction conditions reads
\begin{eqnarray}\label{484}
A_*=A_b,~~~\omega_*=\omega_b,~~~\text{and}~~~C_*=(r_b^2+\omega_b^2)^{-1/2}C_b
\end{eqnarray}
\item[(b)] The extrinsic curvature tensor of the boundary surface must be the same when calculated in both the inner and the outer metrics. 
\end{itemize}

The extrinsic curvature tensor is calculated as follows. Let $n^\mu$ be the outward-pointing unit normal to the surface $r=r_b$, which is to say the unit vector pointing in the direction of increasing $r$. This vector is
\begin{eqnarray}\label{485}
n_{\text{in}}^\mu = B^{-1}\delta_r^\mu,~~n_{\text{out}}^\mu=B_*^{-1}\delta_r^\mu
\end{eqnarray}
in the inner and outer regions respectively. 

If $\tau^\mu$, $\vartheta^\mu$ and $\zeta^\mu$ are unit vectors point in the directions of increasing $t$, $\theta$ and $z$ respectively, then the extrinsic curvature tensor $K_{AB}$ is a tensor on the 3-dimensional boundary surface with components
\begin{eqnarray}\label{486}
K_{tt}
&=& \tau^\mu\tau^\nu\nabla_{\mu}n_{\nu},~K_{\theta\theta} =\vartheta^\mu\vartheta^\nu\nabla_{\mu}n_\nu,~K_{zz}=\zeta^\mu\zeta^\nu\nabla_\mu n_\nu\nonumber\\
\end{eqnarray}
and
\begin{eqnarray}\label{487}
K_{t\theta}
&=&K_{\theta t} =\frac{1}{2}(\tau^\mu\vartheta^\nu +\vartheta^\mu\tau^\nu)\nabla_\mu n_\nu
\end{eqnarray}
\begin{eqnarray}
K_{tz}=K_{zt}=\frac{1}{2}(\tau^\mu\zeta^\nu +\zeta^\mu\tau^\nu)\nabla_\mu n_\nu
\end{eqnarray}
\begin{eqnarray}
K_{\theta z}=K_{z\theta}=\frac{1}{2}(\vartheta^\mu\zeta^\nu+\zeta^\mu\vartheta^\nu)\nabla_\mu n_\nu
\end{eqnarray}
where $\nabla_a$ is the covariant derivative corresponding to the metric, which of course will be different in the inner and outer regions. Note however that
\begin{eqnarray}\label{488}
\nabla_\mu n_\nu = \partial_\mu n_\nu -\Gamma_{\mu\nu}^\lambda n_\lambda
\end{eqnarray}
where $\partial_\mu$ denotes $\partial/\partial x^\mu$ and $\Gamma_{\mu\nu}^\lambda$ is the Christoffel symbol. 

Since $B$ and $B_*$ in our problem do not depend on $t,~\theta$ or $z$, and $\Gamma_{\mu\nu}^\lambda$ is symmetric in the indices $\mu$ and $\nu$ (meaning that $\Gamma_{\nu\mu}^\lambda=\Gamma_{\mu\nu}^\lambda$) the components of the extrinsic curvature tensor can be written in the more compact form
\begin{eqnarray}\label{489}
K_{AB}
&=& -\Gamma_{AB}^\lambda n_\lambda
\end{eqnarray}
where the indices $A$ and $B$ range over the set $\left\{t,\theta,z\right\}$. So (given that $n_\lambda = g_{\lambda\mu} n^\mu = -B\delta_\lambda^r$ in both regions)
\begin{eqnarray}\label{490}
K_{AB}^{\text{in}}=(B_b~\Gamma_{AB}^r)^{\text{in}}~~~\text{and}~~~K_{AB}^{\text{in}}=(B_*~\Gamma_{AB}^r)^{\text{out}}
\end{eqnarray}
and the second set of junction conditions requires that $(B_b~\Gamma_{AB}^r)^{\text{in}}=(B_*~\Gamma_{AB}^r)^{\text{out}}$ on the boundary surface.

To write down the second set of junction conditions explicitly, note that in the outer metric the only non-zero components of the Christoffel symbols are
\begin{eqnarray*}
\Gamma_{\theta r}^\theta =\Gamma_{r\theta}^\theta =\frac{1}{r},~~~\Gamma_{\theta r}^z=\Gamma_{r\theta}^z=\frac{\omega_*C_*^2}{r}
\end{eqnarray*}
\begin{eqnarray}\label{491}
\Gamma_{\theta\theta}^r=-\frac{C_*^2}{B_*^2}r
\end{eqnarray}
According to the results in Chapter 8, the only non-zero components of $\Gamma_{AB}^r$ in the inner region are
\begin{eqnarray}\label{492}
\Gamma_{tt}^r
&=&\frac{AA'}{B^2},~~~\Gamma_{\theta\theta}^r=-\frac{CC'}{B^2},~~~\Gamma_{\theta z}^r=\Gamma_{z\theta}^r=\frac{\omega'}{2B^2}~~~\text{and}~~~\nonumber\\
&&\Gamma_{zz}^r=-\frac{r+\omega\omega'}{B^2C^2} +\frac{r^2+\omega^2}{B^2C^3}
\end{eqnarray}

Altogether, therefore, the second set of junction conditions entails that
\begin{eqnarray}\label{493}
\frac{A_bA_b'}{B_b^2}&=& 0,~~~\frac{C_bC_b'}{B_b^2}=\frac{C_*^2}{B_*^2}r_b,~~~\frac{\omega_b'}{2B_b^2}=0,~~~\text{and}~~~ \nonumber\\
&&-\frac{r_b+\omega_b\omega_b'}{B_b^2C_b^2} +\frac{(r_b^2+\omega_b^2)C_b'}{B_b^2C_b^3} =0
\end{eqnarray}
or equivalently, given that $\omega_*=\omega_b$ and $C_*=(r_b^2+\omega_b^2)^{-1/2}C_b$
\begin{eqnarray}\label{494}
A'_b=0,~~~B_*=\left(\frac{B_b^2}{C_b'}\frac{r_bC_b}{r_b^2+\omega_b^2}\right)^{1/2},~~~\omega_b'=0
\end{eqnarray}
and
\begin{eqnarray}
C_b'=\frac{r_b}{r_b^2+\omega_b^2}C_b.
\end{eqnarray}
The second and last of these conditions can be used to show that
\begin{eqnarray}\label{495}
B_*=B_b
\end{eqnarray}

Unfortunately, however, the condition $A_b^{'}=0$ can never be satisfied, for the same reason that $A$ can not remain finite in solutions extending from $r=0$ to $r=\infty$. If we define $U=rA'/A$ then, as we seen previously, we have $U=0$ at $r=0$, but $U$ can never return to 0 if $K_s\neq 0$. 

So, there are no solutions with a finite radius that can satisfy the junction conditions at the boundary radius $r=r_b$.

\begin{center}
\textbf{Acknowledgment}
\end{center}
This research was fully funded by Graduate Research Scholarship from Universiti Brunei Darussalam (GRS UBD) for MH. This support is greatly appreciated.
\\~~\\

\end{document}